# Cell types and ontologies of the Human Cell Atlas

**Authors**: David Osumi-Sutherland[1], Chuan Xu[2], Maria Keays[2], Peter V. Kharchenko[3], Aviv Regev[4], Ed Lein[5], Sarah A. Teichmann[2,6]

**Affiliations**:
1: EMBL-European Bioinformatics Institute, Wellcome Genome Campus, Hinxton, Cambridge CB10 1SD, UK
2: Wellcome Sanger Institute, Wellcome Genome Campus, Hinxton, Cambridge CB10 1SA, UK
3: Department of Biomedical Informatics, Harvard Medical School, Boston, Massachusetts 02115, USA
4: Genentech, 1 DNA Way, South San Francisco, California 94080, USA
5: Allen Institute for Brain Science, Seattle, Washington 98109, USA
6: Cavendish Laboratory, University of Cambridge, JJ Thomson Ave, Cambridge CB3 0HE, UK
Correspondence should be addressed to S.A.T.

**Abstract**
Massive single-cell profiling efforts have accelerated our discovery of the cellular composition of the human body, while at the same time raising the need to formalise this new knowledge. Here, we review current cell ontology efforts to harmonise and integrate different sources of annotations of cell types and states. We illustrate with examples how a unified ontology can consolidate and advance our understanding of cell types across scientific communities and biological domains.

**Main**
With collaboration of over 2,000 scientists across more than 1,000 institutes from 76 countries to date, the Human Cell Atlas (HCA) has generated comprehensive molecular profiles of tens of millions of single cells across 18 different organs and systems, which in turn, are advancing our understanding of the definition of cell types and states[1,2]. The technological revolution in single-cell and spatial genomics is rapidly expanding the compendium of known cell types, and accelerating discoveries of a large variety of novel cell populations.

For instance, in the field of immunology, where well-established cell types have been historically recognised and well categorised, the numbers of cell subsets identified from single-cell genomics (reflecting both discrete types and specific states) exceeded expectations, particularly with respect to the diversities of cell states derived from developmental dynamics[3], tissue-resident phenotypes[4] and activation states[5]. For example, in the maternal-fetal interface, three decidual natural killer cell populations were identified with varying levels of immunoregulatory properties in order to properly modulate trophoblast invasions[6].

This wide variety of cell types and gene programmes is also apparent in the central and peripheral nervous systems. The number of known cell types previously characterised in the mammalian brain has been vastly exceeded by those uncovered by cell atlasing of mouse and human brains, where over a hundred cell types have been described in just a single region of the neocortex[7]. Recent single-cell studies across mammalian brains highlight the cellular diversity due to species-specific adaptations in the neocortical excitatory neurons[7-9], including increased cell type diversity associated with human cortical expansion[8], and molecular and physiological specialisations of the spinal cord-projecting Betz cells in humans and non-human primates[9]. Similar dramatic increase in known diversity is now being uncovered in the peripheral nervous system, as has been the case in the enteric nervous system[10, 11].

This incredible progress takes us closer to the answer to a general question motivating the HCA project: what is the complete cellular makeup of the human body? Annotating cells and gene programmes is crucial in order to address this question, and more fundamentally, to fully leverage such data for biological discovery and knowledge. This can only be achieved by naming the entities we study in a consolidated way, such that findings can be related between studies and one study can build on findings from multiple previous ones as knowledge is accrued and expanded. However, most annotations of single-cell genomics datasets to date have used uncontrolled free text for cell type names, making cross-searching of annotations across separate datasets challenging and unreliable. In some cases, with a naming scheme absent, cells are described merely by a subset of their molecular characteristics, and thus can be hard to match between studies.

To fully answer the question of what the cellular composition of the human body is, there is an urgent need to put new discoveries from the HCA into the context of classical cell biology and anatomy, as well as developmental biology, neurobiology and clinical biology. Cell ontologies are a tremendously powerful way of formalising such knowledge, which in turn open up opportunities for quantitative scientific interrogation of the HCA data in new and exciting ways.

Efforts to implement controlled vocabularies for cell type annotations are already underway, with the ontologies of cell types including those from Cell Ontology[12] and the *Drosophila* Anatomy Ontology[13] being increasingly widely used to annotate single-cell transcriptomic data. This makes cell annotations cross-searchable and accessible to biologists in general, and more importantly, relates annotated data back to hard-earned legacy knowledge, classical terminologies, and the accompanying understanding of cell types, anatomies and development. These ontologies incorporate simple categorical assertions about the properties of cell types including their relationships to anatomy, lineage, function and gene expression (where that is critical to defining a cell type). While these simple categorical definitions may not reflect the often-continuous and variable nature of biology at the single-cell level, by making annotated single-cell data findable and integrable, they facilitate cross-dataset analyses that can provide a more nuanced and statistical view of cell types, their classification, and their properties based on analyses of similarities between thousands of individual cells. Cell Ontologies have now begun to supplement these categorical definitions of cell types with classifications and links to reference data directly derived from these analyses.

In this Perspective, we discuss the essential utility and key parts of cell ontologies, review the state of current cell ontologies, and conclude with the ongoing efforts and how they can be applied to a variety of use cases over the coming years.

**Using cell ontology for knowledge integration and mining**

Biomedical ontologies originated in simple controlled vocabularies developed to supplement or replace the free text metadata in databases, clinical records and medical billing systems[14]. Standardising the text used to record, for example, diseases, gene functions, anatomical structures, and cell types within and between databases makes it possible to reliably search and group records referring to the same entities (diseases, cell types, etc.). However, it is not sufficient for searching and grouping records with closely related contents. For example, a user searching a database for records relating to macrophages or liver sinusoid would not find records for Kupffer cells unless the data structures driving the search had some meaningful ways to relate the terms 'macrophage', 'Kupffer cell' and 'liver sinusoid'.

Cell ontologies, conversely, provide mechanisms for this integration, allowing us to record a 'Kupffer cell' as a type of macrophage located in the liver sinusoid and then to enrich search results to take advantage of the classification and location relationships (Fig. 1a). Because the precise terms used for cell types, anatomical structures and diseases often vary greatly across sources, biomedical ontologies, including the cell ontology, typically use a bipartite system of universally resolvable IDs to link official labels and synonyms[15]. Critically, using resolvable IDs to denote ontology terms allows associated metadata (labels, synonyms, descriptions and references) and their relationships to evolve over time with no cost for the databases and records that use them.

Ontologies can also serve to link and integrate heterogeneous data types across multiple modalities all related to the same cell type. For example, Virtual Fly Brain[16, 17] and the Fly Cell Atlas (Stein Aerts and Norbert Perrimon, personal communication; https://flycellatlas.org) use the same ontology terms to annotate 3D images of neurons (>70,000 images), connectomics data (>3.5 million pairwise connections), and single-cell transcriptomics data (~600,000 cells). Similarly, Cell Ontology terms, classifications and relationships are also increasingly used to define and classify terms in the Gene Ontology[18] (>750 terms) and in widely-used ontologies of phenotypes (730 terms in the Human Phenotype Ontology[19]) and disease (>3,000 terms in the Mondo disease ontology[20]). These links make it possible to combine single-cell, phenotype and disease data relating to the same cell types. With the advent of large-scale single-cell transcriptomic atlasing, community-driven nomenclature and ontology building projects have emerged and are coordinating with existing ontology building efforts (e.g., HCA Biological Networks[2], HuBMAP[21], BRAIN Initiative Cell Census Network (BICCN)[22], Cell Annotation Platform (http://celltype.info)).

This is already impacting our ability to organise our knowledge of cell types for comparisons of datasets across individual laboratories, and notably, for effectively interpreting health and disease using the knowledge from both classical histopathology and single-cell genomics. For instance, ontological distinctions between fetal and mature cells in the kidney are mirrored by differences in their molecular signatures, which are critical to understanding the divergent origins of pediatric and adult kidney cancers, respectively[23]. Similarly, consistently annotated datasets allowed cross-tissue meta-analyses for COVID-19 that identified specialised nasal epithelial cells enriched for expression of SARS-CoV-2 entry factors[24], identified covariates such as age, sex and smoking status associated with the entry factor expression in lung and airway cells[25], and compared cells in COVID-19 tissues from patient autopsies to healthy and other disease conditions[26], again highlighting the necessity and utility of establishing agreed-upon ontological classifications.

**Considerations in the classification of human cell types**

Biologists have long recognised that the natural world lends itself to hierarchical systems of classification, which capture the underlying hierarchical processes driving biology such as the phylogenetic classification of species by morphological and molecular observations. Several such processes can also drive cell diversifications including ontogeny (cell differentiation), morphogenesis (often driven by continuous gradients), and the dual impact of a cell's differentiation history and tissue context. All of these are imprinted in a cell's molecular properties, and many can be captured by hierarchical representations. Thus, cell types can be classified and categorised in ever-increasing levels of resolution reflecting these underlying processes (even when the process is not explicitly known), from a broad cell type like an endothelial cell, through more specialised types like a liver sinusoidal endothelial cell (LSEC), down to highly specialised types found in specific locations such as a periportal LSEC. As with a species' taxonomy, there are various kinds of observations informing the ultimate classification, and these different types of information are often used in concert to arrive at a particular cell type definition.

Sources of information contributing to a cell type categorisation can include morphological features, gene expression profiles, anatomical locations within an organism or tissue, developmental origins, and functional profiles. Ontologies are able to capture all these kinds of information about a cell type. Take anatomical locations as an example, the Cell Ontology[12] imports information about anatomical structures and features from the Uber-anatomy Ontology (Uberon)[27] via ontological relations such as 'part of' and 'located in'. As a result, the Cell Ontology definition of an LSEC includes the relation "'part of' some 'hepatic sinusoid'", which indicates that the cell forms part of the structure of the hepatic sinusoid as defined in Uberon. In an anatomically higher hierarchy, the definition of hepatic sinusoid involves relations to the liver lobule and the liver overall, which is in turn defined by its structure, location and physiological role in the body. The LSEC is hence hierarchically defined relative to the whole organism down to its individual position in the specific tissue where it is found (Fig. 1a). Furthermore, since the Cell Ontology classifies cell types hierarchically from generic cell types down to more specialised types, an LSEC is also defined as a descendent of the general endothelial cell class in the Cell Ontology. The main LSEC class (officially 'endothelial cell of hepatic sinusoid') has its own descendent classes, representing further specialisations of LSECs: 'endothelial cell of periportal hepatic sinusoid' and 'endothelial cell of pericentral hepatic sinusoid'.

Cell ontologies also represent developmental lineages and, to a more limited extent, cell states such as activation, cycling, reshaping and stressing (Fig. 1b) - either directly or through extensions to existing annotations. For example, in addition to a general term for a microglial cell, the Cell Ontology also defines, as descendants of this class, immature and mature microglial cells, which have the same characteristics as the general class, as well as specific attributes which set them apart as either immature ("microglial cell with a ramified morphology") or mature ("microglial cell with an amoeboid morphology that is capable of cytokine production and antigen presentation"). Cell-cycle states, on the other hand, can be represented in the annotation system by combining a Cell Ontology term with a term from the Gene Ontology Cell Cycle Phase terms.

Developmental samples or actively regenerating tissues present particular challenges to cell ontology development and cell type annotations, as a plethora of intermediate states and continuous branching lineages can be partitioned in different ways and may not lend themselves well to discrete labels. In such a setting, the annotation needs to emphasize the relative ordering of states, or their positions on a continuous differentiation path. There are also striking examples of developmental convergence (developmental homoplasy). Somatosensory neurons, for example, can be equivalently derived from neural crest or sensory placodes[28]. Similarly, dermal fibroblasts in different parts of the trunk or face are derived from distinct sources, despite the overall molecular and phenotypic likeness[29]. Nevertheless, cell ontologies record gross lineage relationships, with limited temporal resolution between developing/progenitor and mature cell types using specific relations where these relationships are stereotyped and consistent. To date, the Cell Ontology records lineage and differentiation relationships for more than 1,900 cell types, connecting developing cell types to developing tissues and stages via links to Uberon. Greater temporal resolution can be achieved in annotation by combining terms specifying this context with Cell Ontology terms at annotation time.

Historically, different fields in biology have focused on different aspects of cells to drive their naming. For example, many immune cells have been classified according to which cell surface protein(s) they express[30-37], whereas cells of the nervous system have been named according to a combination of features including morphologies, physiologies, connectivities and the roles they play in the neuronal circuitry[38]. In some systems, such as the retina[39], there is strong evidence that such features all map to a single ontology, but in other cases different features could in principle lead to different classifications. Ontologies attempt to capture all terms that are used by different scientific communities to refer to the same cell type, as well as alternative names that may not be commonly used. Currently, cell types and states can be elucidated from single-cell genomic data based on transcriptomic, epigenetic and proteomic expression profiles, using different software such as SCCAF[40]. While these inferences are unbiased, it is important to reconcile them with conventional biological understandings and terminologies.

**Current state of ontologies**
The Cell Ontology and Uberon are both species-neutral ontologies with a strong focus on mammalian anatomies and cell types, as well as standard mechanisms for recording the species applicability of terms. The Cell Ontology has 2,311 terms covering all major cell types. The granularity of this coverage is variable, with most coverage currently for the immune system (>500 cell types). Uberon defines over 14,000 types of anatomical structures and records many types of relationships between them. Practically, Cell Ontology and Uberon are tightly integrated with each other: approximately 1,900 cell types in Cell Ontology are linked by part relationships to the anatomical structures defined in Uberon. Further combining the Cell Ontology with the established categories and markers used in the National Health Service (NHS) for Pathology (through consultation with NHS pathologists and clinicians listed in the Acknowledgements), and with the newly discovered cell populations from HCA data, we are able to cover major organs and cell types in the human body (Table 1).

The human applicable components of Cell Ontology and Uberon are under active development as part of multiple collaborative efforts. Terms are being added in a coordinated fashion to both ontology platforms in response to the annotation needs of atlasing projects including HCA's Data Coordination Platform[2] and Data Portal (https://data.humancellatlas.org), and EBI's Single Cell Expression Atlas[41].

The Brain Data Standards Initiative, part of the NIH BRAIN Initiative Cell Census Network, is extending the Cell Ontology with terms for cortical cell types defined by single-cell transcriptomics, with a current focus on the primary motor cortex of human, marmoset and mouse[9]. This work leverages existing efforts on nomenclature standards[42], but importantly aims to use the quantitative hierarchical cell type classification from single-cell genomics as a data-driven foundation for ontological definitions. Multimodal data about these cell types are integrated at different levels of the hierarchy, including their spatial tissue distributions, morphological and physiological properties (derived from Patch-seq experiments), and axonal projection targets. Ultimately such a data-driven approach may be used across the entire human body, providing a common metric in gene usage to measure similarities and potential common developmental origins across organs.

The CCF/HuBMAP ASCT+B effort[21] is working with a wide community of experts to build tables representing the human anatomy and cell type terminology needed for annotating scRNA-seq datasets, as well as crowdsourcing markers for cell types. All entries in these tables are mapped to the Cell Ontology or Uberon where possible, or turned into term requests where new terms are needed. The implicit part relationships in these tables are validated against Cell Ontology and Uberon with the results of this validation feeding back to improve both the tables and Uberon and Cell Ontology based on the discussion and agreement with experts. The ASCT+B project is building an expert-validated ontological model of the human vasculature that is feeding back hundreds of new terms and relationships into Uberon. One important result of this work will be a curated subset of Cell Ontology and Uberon terms for reliably annotating human scRNA-seq data.

As part of the Cambridge Cell Atlas project (https://www.cambridgecellatlas.org), an effort to make results from single-cell gene expression experiments easily accessible to a broad community of users including clinicians, the Cell Ontology is being enriched and extended based on contributions from pathologists and clinicians. This will introduce human cell types annotated with details of specific clinical markers used to identify them in medical pathology laboratories. This ontology can then be integrated into the search functionality of the Cambridge Cell Atlas platform to enable searching based on for example a marker panel, which brings back gene expression data from cells annotated with those markers.

In summary, the Cell Ontology is continuously evolving with inputs from various projects and perspectives, and meanwhile supplies formalised ontology information back to them (Table 2).

**Applications of a cell ontology**
Cell ontologies provide a single place to look up cell types for the community. Through this, knowledge can be aggregated and standardised in an encyclopaedic sense. First, cross-modal data integration can reinforce or refine the identity of a cell type. For example, the survey on the mammalian neocortex revealed the correspondence of various cellular properties when overlapping imaging, electrophysiology and connectivity with transcriptomic profiles[38]. Second, mining of an ontological classification system can reveal major trends with respect to shared cell types across organ-specific atlases (e.g., immune and endothelial cells) versus specialised types (e.g., goblet cell in the gut and lung), emphasizing the concept of a tissue's function being the collective of its cells operating in a specific 3D organisation in concert.

Importantly, with more single-cell resources employing the cell and anatomy ontologies including but not limited to the Fly Cell Atlas (and cross-species projection via ontology), EBI's Single Cell Expression Atlas and Sanger-EBI Cambridge Cell Atlas, cell ontologies have been linking scientific and medical communities through common nomenclatures and markers for human cell biology, medical pathology and disease. For example, a fine-grained ontological classification of human uterine cells, which was derived from single-cell and spatial transcriptomics, was utilised to interrogate the cellular compositions for endometrial adenocarcinomas[43], revealing the dominance of *SOX9+* epithelial cells which are enriched in the proliferative phase, among which the *SOX9+LGR5+* subpopulation is dramatically expanded in more advanced cancer stages (Fig. 1c).

The application of cell ontologies will be most pertinent in the context of interactive and automated systems for the interpretation and annotation of single-cell genomic datasets. A number of efforts to design such systems are currently under way, including several automated cell annotation projection pipelines[44-47]. For example, as part of the HCA initiative, the Cell Annotation Platform (CAP) aims to provide a general repository for cell annotations of different datasets, in combination with interactive tools for annotating new datasets. For a cell subset of interest, CAP user interfaces will suggest the appropriate ontology terms based on text search, learned synonyms, and eventually molecular signatures themselves. Conversely, the resulting knowledgebase of commonly used annotation terms and associated molecular signatures will provide a useful resource to extend ontologies as well as to train and optimise machine learning models that automate the annotation task. In parallel to these efforts, data-driven ontology development is advancing the community engagement in specific research domains such as NeMO Analytics (for brain, https://nemoanalytics.org) and gEAR (for ear)[48].

**Summary and outlook**
Resolving the cellular makeup of the human body warrants the categorisation of cells in a standardised framework. The Cell Ontology offers one such avenue to consolidating this knowledge in an encyclopaedic manner, with applications from cell and tissue biology all the way to the clinic. Despite potential ambiguities and transitory states a cell may have, each facet of a cell ranging from morphological to molecular features can be taken into account, until a defining status is reached and further recognised by the community.

Many HCA-related resources, such as CAP, have been using the Cell Ontology for *de novo* cell annotation. Cell ontologies also serve other sources of data by retrieving or delivering ontology-level information. We anticipate the synergy between HCA project and Cell Ontology will continue to grow over the coming years and beyond even the completion of HCA, with dimensions of human genetic variation, ageing and disease on the horizon.

In summary, HCA data provide a genomic foundation for human cell ontologies, which will be facilitating further future research, as it is the only way to define a cell type that is universal across the entire body. This will become more pressing and also more clear as the number of HCA studies of individual tissues and organs increases. Importantly, the HCA Biological Networks will provide nucleation points for expert community efforts in order to achieve gold standard consensus cell annotations with the cell ontology terms used for various purposes such as automated annotation. Common cellular phenotypes and developmental origins will become understandable through common gene usage, and functional similarities will be revealed in universal gene patterns. Whole-body consequences of disease will be understandable as well through gene usage differences in cell types throughout the human body. This will thus create opportunities for a new and different kind of quantitative data-driven framework extending and potentially transforming existing ontology efforts.

**Acknowledgements**
We are grateful to Jana Eliasova (scientific illustrator) for support with the figure, to Roser Vento-Tormo for comments on the figure and texts, and to the following clinicians and researchers for information on standard pathology markers for tissues and cells used in the UK National Health Service: Lia Campos, Andrew Dean, Luiza Moore, Neil Sebire, Teresa Brevini, Muzlifah Haniffa, Jing Eugene Kwa, James McCaffrey, and Alexandra Kreins. Research reported in this publication was supported by the Wellcome Trust Grant 108413/A/15/D, the Office Of The Director, National Institutes Of Health of the National Institutes of Health under Award Number OT2OD026682' and grants from the CZI (Chan Zuckerberg Initiative DAF, an advised fund of Silicon Valley Community Foundation).

**Competing interests**
In the last 3 years, S.A.T. has been remunerated for consulting and SAB membership by Foresite Labs, GlaxoSmithKline, Biogen, Roche and Genentech. A.R. is a founder of and equity holder in Celsius Therapeutics, an equity holder in Immunitas Therapeutics, and was a scientific advisory board member for ThermoFisher Scientific, Syros Pharmaceuticals and Neogene Therapeutics until August 1, 2020. From August 1, 2020, A.R. is an employee of Genentech. A.R. is a named inventor on several patents and patent applications filed by the Broad Institute in the area of single cell and spatial genomics.

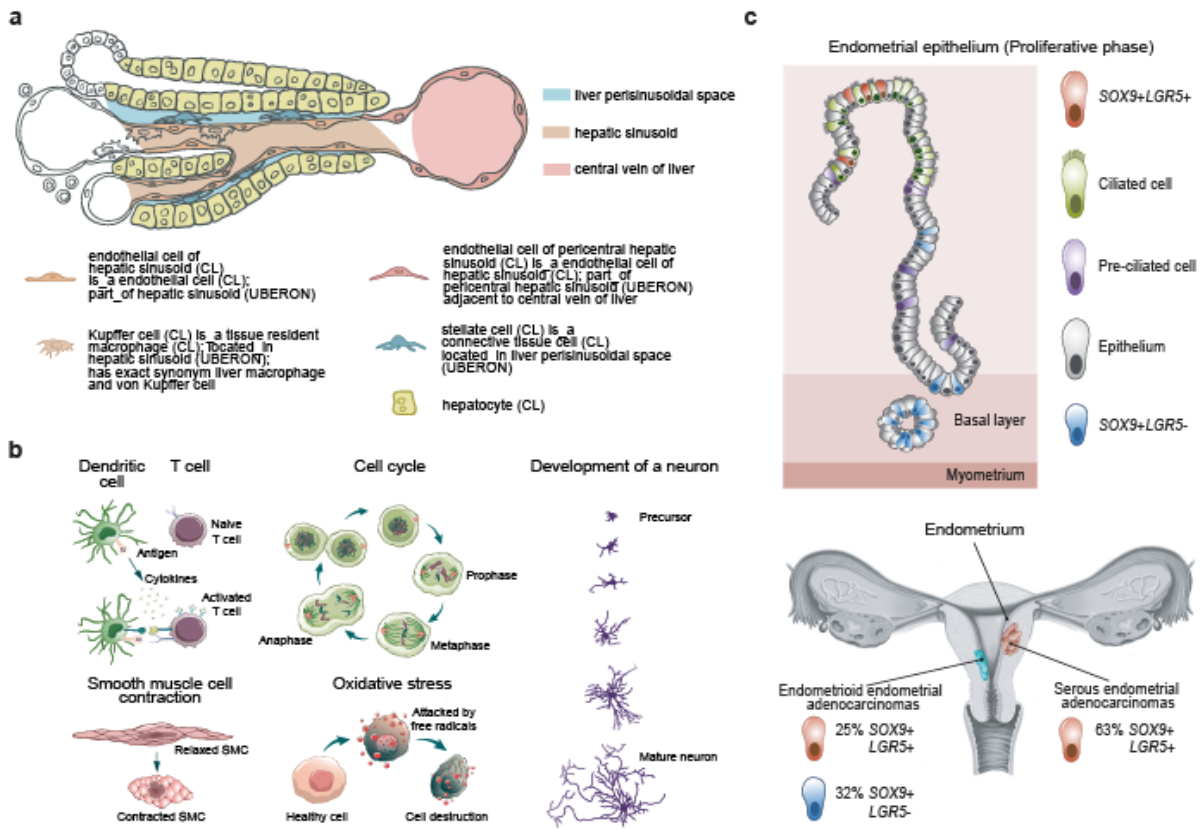

**Fig. 1: Cell ontology as a way of linking human cell types with anatomy, cell state transition, and disease. a**, As well as providing IDs, synonyms and descriptions of cell types, ontologies allow us to record the classification of cell types in the anatomical context. Here, the Cell Ontology has terms for a variety of cell types associated with the hepatic sinusoid (UBERON: 0001281). The classification of these cell types allows them to be grouped with other cells from the same location (e.g., Kupffer cells (CL: 0000091) can be grouped with other tissue-resident macrophages or with cells of the hepatic sinusoid). **b**, Ontologies can be used to encode the cell transitions through diverse types of cell states such as T cell activation following antigen recognition, cell cycling, neuron development and maturation, smooth muscle cell contraction and relaxation, and cell destruction after the oxidative stress. **c**, Schematic delineating the cell types present in human endometrial epithelium summarised from Garcia-Alonso et al., 2021[43]. Cell ontologies link single-cell genomics with endometrial cancer, and reveal how the proliferative phase-enriched *SOX9+* epithelial population including its *LGR5+* and *LGR5-* subpopulations can vary in compositions along different stages of the endometrial cancer.

**Table 1 Current status of cell type enumerations in NHS Pathology, Cell Ontology and HCA data.** Summary of cell type numbers from NHS Cellular & Molecular Pathology Royal College of Pathology, Cell Ontology, and HCA data.

| Tissue | No. cell types (NHS Pathology) | No. cell types (Cell Ontology version: 2021-04-22) | No. cell types as per HCA Ref | HCA Ref |
|---|---|---|---|---|
| Kidney | 9 | 126 | 33 (mature)/44 (fetal) | Stewart et al., 2019[49] |
| Lymph node | 8 | 3 | 19 | James et al., 2020[50] |
| Small and large intestine | 9 | 90 | 132 | Elmentaite et al., 2021[11] |
| Lung | 8 | 27 | 21; 58 | Vieira Braga et al., 2019[51]; Travaglini et al., 2020[52] |
| Liver | 12 | 15 | 21; 39 | Ramachandran et al., 2019[53]; Aizarani et al., 2019[54] |
| Muscle | 8 | 31 | 19 | Litviňuková et al., 2020[55] |
| Esophagus | 2 | 11 | 18 | Madissoon et al., 2019[4] |
| Heart | In progress | 54 | 67 | Litviňuková et al., 2020[55] |
| Thymus | In progress | 55 | 44 | Park et al., 2020[56] |
| Brain | 10 | 108 | 127 | Bakken et al., 2020[9] |
| Bone marrow and blood | In progress | 515 | 48 | HCA Data Portal |
| Skin | In progress | 71 | 34 | Reynolds et al., 2021[57] |
| Endometrium and decidua | In progress | 5 | 14; 11 | Garcia-Alonso et al., 2021[43]; Vento-Tormo et al., 2018[6] |
| Placenta | 6 | 10 | 5 | Vento-Tormo et al., 2018[6] |

**Table 2 Projects using and contributing to the Cell Ontology (CL).**

| Project | Description | CL Use | URL |
|---|---|---|---|
| Cell Annotation Platform | An open annotation platform for scRNA-seq data | Uses CL and free text for cell type annotation | http://celltype.info |
| EBI Single Cell Expression Atlas & Cambridge Cell Atlas | Open public repository for exploration of single cell gene expression data | Uses CL to annotate samples and cell types in tertiary analysis | https://www.ebi.ac.uk/gxa/sc and https://www.cambridgecellatlas.org |
| HCA/DCP | Community generated, multi-omic, open data processed by standardized pipelines | Uses CL to annotate samples and cell types in tertiary analysis | https://data.humancellatlas.org |
| HuBMAP/CCF ASCT+B tables | Expert curated tables of human cell types, their markers and anatomical context | Maps all cell types to CL | https://hubmapconsortium.github.io/ccf-asct-reporter |
| cellxgene | An open annotation platform requiring annotation with ontology terms | Uses CL to annotate samples and cell types in tertiary analysis | https://chanzuckerberg.github.io/cellxgene |
| Tabula Muris | Curated whole mouse scRNA-seq atlas | Uses CL to annotate gross cell types, extending definitions with free text and markers | https://tabula-muris.ds.czbiohub.org |
| Monarch Initiative | A resource building ontologies of phenotypes and disease and using these to build an integrated collection of phenotype/disease to gene/variant associations | Defines cellular phenotypes and diseases | https://monarchinitiative.org |
| Gene Ontology | The world's largest source of information on the function and location of gene products | Defines cell type-specific organelles and biological processes | http://geneontology.org |